\documentclass[runningheads]{llncs}
\usepackage[T1]{fontenc}
\usepackage{cite}
\usepackage{amssymb, amsmath, latexsym}
\usepackage{algorithm}
\usepackage{algpseudocode}
\usepackage{booktabs}
\usepackage{subcaption}
\usepackage{bm}
\usepackage{hyperref}
\usepackage{orcidlink}

\makeatletter

\usepackage[font=scriptsize]{caption}

\usepackage{wrapfig}
\usepackage{array}
\newcolumntype{P}[1]{>{\centering\arraybackslash}p{#1}}

\usepackage{graphicx}

\begin{document}
\title{Suppressing Poisoning Attacks on Federated Learning for Medical Imaging}

\author{Naif Alkhunaizi\orcidlink{0000-0002-7093-5034
} \and
Dmitry Kamzolov\orcidlink{0000-0001-8488-9692} \and 
Martin Tak\'a\v{c} \orcidlink{0000-0001-7455-2025}  \and
Karthik Nandakumar\orcidlink{0000-0002-6274-9725
} }

\authorrunning{N. Alkhunaizi et al.}

\institute{Mohamed bin Zayed University of Artificial Intelligence \\
\email{\{naif.alkhunaizi, kamzolov.dmitry, martin.takac, karthik.nandakumar\}@mbzuai.ac.ae}}

\maketitle          
\begin{abstract}
Collaboration among multiple data-owning entities (e.g., hospitals) can accelerate the training process and yield better machine learning models due to the availability and diversity of data. However, privacy concerns make it challenging to exchange data while preserving confidentiality. Federated Learning (FL) is a promising solution that enables collaborative training through exchange of model parameters instead of raw data. However, most existing FL solutions work under the assumption that participating clients are \emph{honest} and thus can fail against poisoning attacks from malicious parties, whose goal is to deteriorate the global model performance. In this work, we propose a robust aggregation rule called Distance-based Outlier Suppression (DOS) that is resilient to byzantine failures. The proposed method computes the distance between local parameter updates of different clients and obtains an outlier score for each client using Copula-based Outlier Detection (COPOD). The resulting outlier scores are converted into normalized weights using a softmax function, and a weighted average of the local parameters is used for updating the global model. DOS aggregation can effectively suppress parameter updates from malicious clients without the need for any hyperparameter selection, even when the data distributions are heterogeneous. Evaluation on two medical imaging datasets (CheXpert and HAM10000) demonstrates the higher robustness of DOS method against a variety of poisoning attacks in comparison to other state-of-the-art methods. The code can be found here \url{https://github.com/Naiftt/SPAFD}.

\keywords{Federated learning  \and Parameter aggregation \and Malicious clients \and Outlier suppression}
\end{abstract}

\section{Introduction}

\begin{figure}[t]
\centerline{\includegraphics[scale = 0.4]{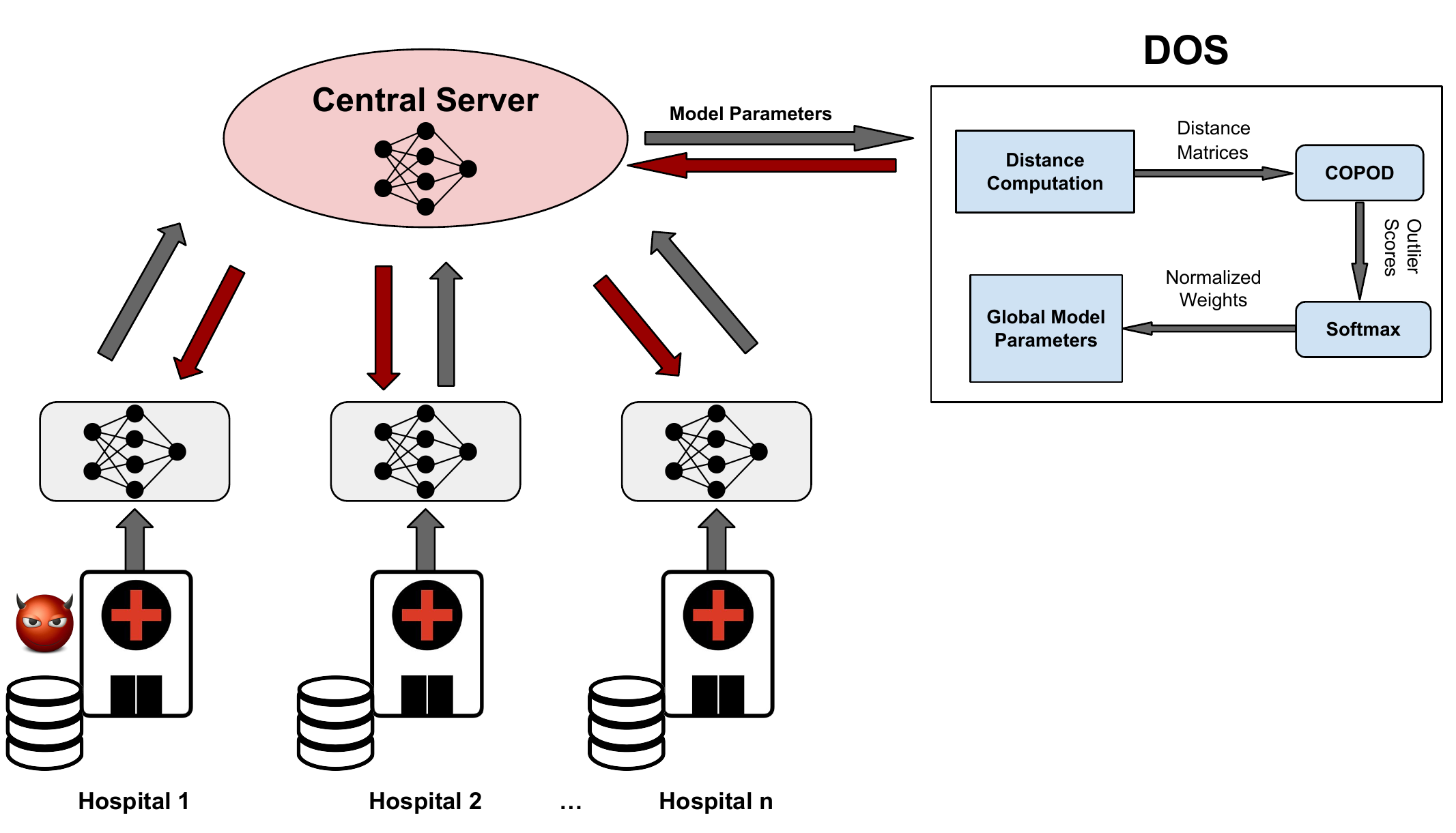}}
\caption{Federated Learning (FL) in the presence of a malicious client. Distance-based Outlier Suppression (DOS) aggregation rule (right) can be applied to robustly aggregate local parameter updates.}  \label{figDOD}
\end{figure}

Medical institutions often seek to leverage their data to build deep learning models that work jointly with physicians as decision support systems for predicting different diseases \cite{liu2020clinical, esteva2017dermatologist}. However, limited accessibility to data from external healthcare institutions can impede the learning process \cite{vanPanhuisDataBarriers2014} or introduce bias towards the local data \cite{kaushal2020health}. Hence, collaborative learning is vital to expand data availability and maximize model performance. Federated Learning (FL) \cite{li2020federated} provides a paradigm where multiple institutions/devices can use their data to train local models and exchange their model parameters regularly with a central server. The server aggregates the local model parameters to update a global model, which is shared back to the clients. This allows all parties to preserve their data locally and offers better results while maintaining data privacy. Recent works have underscored the importance of FL in medical imaging \cite{ShellerFLInMedicineNSR2020,DayanFLCOVID-19NM2021}.

Typically, FL algorithms (e.g., FedAvg \cite{mcmahan2017communication}) operate under the assumption that clients participate in the protocol honestly and their data distributions are identical. Techniques proposed to tackle non-iid data \cite{li2020federated} cannot handle malicious parties that attempt to deteriorate the global model performance. Real-world FL is susceptible to a variety of targeted/untargeted attacks \cite{zhou2021deep,fu2019attack} and byzantine faults \cite{lamport2019byzantine}. Though many robust aggregation rules have been proposed for byzantine-tolerant FL \cite{guerraoui2018hidden,yin2018byzantine,blanchard2017machine}, such methods often require a careful choice of hyperparameters (e.g., prior knowledge of proportion of malicious clients) and are vulnerable to more sophisticated attacks \cite{fang2020local}. 

The core contribution of this work is a novel robust aggregation rule called Distance-based Outlier Suppression (DOS) for byzantine-tolerant FL. The proposed framework (see Figure \ref{figDOD}) enables the central server to suppress local parameter updates from malicious clients during aggregation. The DOS framework is enabled by computing distances (Euclidean and cosine) between local parameter updates of different clients and detecting outliers in this distance space. A state-of-the-art, parameter-free outlier detection algorithm called COPOD \cite{li2020copod} is applied to compute the outlier score for each client based on the above distances. Finally, the outlier scores are mapped to client-specific weights, and weighted average of local parameter updates is utilized for global model update. The proposed DOS aggregation rule demonstrates high robustness against diverse poisoning attacks under both iid (CheXpert) and non-iid (HAM10000) settings.

\section{Related Work}

Most FL algorithms aim to minimize the following loss function: 
\begin{equation}
\label{eqn:FLobjective}
\textstyle{\min_{\theta \in \mathbb{R}^d}} \left\lbrace \mathbf{F}(\theta) := \textstyle{\sum_{i=1}^{n} \alpha_i \mathbf{F}_i(\theta)}\right\rbrace,
\end{equation}

\noindent where $n$ is the number of clients, $F_i$ is the loss function for the $i^{th}$ client, and $\alpha_i$ is a value between 0 and 1 that weights the contribution of the $i^{th}$ client. Popular aggregation rules such as FedAvg \cite{mcmahan2017communication} either assign equal weights to all clients ($\alpha_i=1/n, \forall~i=1,2,\ldots,n$) or assign weights to the clients based on the relative size of their training sets. Such schemes have shown effective outcomes under the assumption of honest clients and iid data distributions. FedProx \cite{li2020federated} was introduced to tackle heterogeneity and non-iid data across clients. Progressive Fourier Aggregation (PFA) was introduced in \cite{chen2021personalized} to improve stability by preventing an abrupt drop in accuracy. While these methods promise convergence, they do not consider noisy parameters \cite{wei2022federated} or malicious parties that attempt to hinder learning or cause it to converge to poor local minima.

Since FL requires regular communication between the server and clients, it is susceptible to random network errors that can deliver abnormal parameters to the server or malicious parties that attempt to corrupt the learning process. Such failures are defined as Byzantine failures \cite{lamport2019byzantine} due to the erratic behavior of adversarial parties. In such settings, there are two types of attacks: (i) targeted attacks that cause the global model to misclassify a selected class (e.g., backdoor attacks \cite{bagdasaryan2020backdoor}, model poisoning attacks \cite{bhagoji2018model, bhagoji2019analyzing}, and data poisoning attacks \cite{munoz2017towards}) and (ii) untargeted attacks that harm the model's performance across all classes \cite{ li2020federated}. Moreover, there are techniques designed to make the attack more stealthy \cite{zhou2021deep}. Other potential attacks include label-flipping, where the model is trained on incorrect labels \cite{fu2019attack}, and crafted model attacks \cite{fang2020local} that formulate the attack as an optimization problem. In this paper, we mainly focus on untargeted attacks since they can cause more damage to the learning process. For example, our experiments for FedAvg showed that flipping a tiny subset of the class does not affect the accuracy of the global model, even though FedAvg is not considered a byzantine robust aggregation method. Nevertheless, label-flipping will affect the accuracy of the individual targeted class.

Robust aggregation rules that can defend against Byzantine failures have been proposed. Krum \cite{blanchard2017machine} calculates a score for each client based on the distance between its parameter update and mean parameter update of all clients and selects the client with the lowest score. Coordinate-Wise-Median \cite{yin2018byzantine} calculates the median because it is less sensitive to outliers as opposed to the mean. Trimmed Mean \cite{yin2018byzantine} sorts the parameter values and discards a fraction of each coordinate's smallest and largest values. An enhancement to the above aggregation rules known as Bulyan was proposed in \cite{guerraoui2018hidden} to further improve the robustness. However, these methods require hyperparameter selection (e.g., proportion of outliers to be discarded) and are susceptible to crafted model attacks \cite{fang2020local}.

\begin{algorithm}[t]

\caption{Distance-based Outlier Suppression (DOS)}\label{alg:cap}
\begin{algorithmic}[1]
\Require Initialize parameter vector $\theta^0 \in  \mathbb{R}^d$ for global model
        
\For{$t=0,1,\ldots,T-1$}
        \State \underline{Server}: Server broadcasts $\theta^{t}$ to all clients
        \For{all $i \in [n]$ in parallel}
        \State \underline{Client i}: Learn local parameters $\theta_i^{t+1}$ using suitable local optimizer. 
        \State Send $\hat{\theta}_i^{t+1}$ to server. Note that $\hat{\theta}_i^{t+1} \neq \theta_i^{t+1}$ for malicious clients.
        \EndFor
    \State \underline{Server}: \textbf{DOS} aggregation rule:
    \\ $\qquad$ Compute Euclidean and cosine distance matrices $M_{E}$ and $M_{C}$, respectively
    \\ $\qquad$ Compute outlier scores $\mathbf{r}_{E} = COPOD(M_{E})$, $\mathbf{r}_{C} = COPOD(M_{C})$
    \\ $\qquad$ Compute average outlier score $\mathbf{r} = (\mathbf{r}_{E} + \bm{r}_{C})/2$
    \\ $\qquad$ Compute normalized client weights as $w_{i} = \tfrac{\exp(-r_{i})}{\sum_{j=1}^n exp(-r_{j})}$
    \\ $\qquad$ Update global model based on weighted average of local parameters
    $$
        \theta^{t+1}= \textstyle{\sum}_{i=1}^n w_i \hat{\theta}_{i}^{t+1}
    $$
\EndFor
\end{algorithmic}
\end{algorithm}

\section{Proposed Method}

We propose Distance-based Outlier Suppression (\textbf{DOS}), a robust FL aggregation rule (see Algorithm \ref{alg:cap}) that can defend against different untargeted poisoning attacks on FL as long as the proportion of clients experiencing byzantine-failures is less than $50$\%. Suppose that there are $n$ clients in the FL setup, and their goal is to learn the global model parameters $\theta$ that minimizes the objective function in equation (\ref{eqn:FLobjective}). Note that a proportion $p$ of these clients ($p < 50\%$)  can be malicious and may not share the same objective as the other honest clients. Similar to FedAvg, the global model parameters are initialized to $\theta^0$. In each of the $T$ communication rounds, the current global model parameters $\theta^t$ are broadcast to all clients. Each client $i \in [n]$ learns the local model parameters $\theta_i^{t+1}$ based on their local data using a suitable optimizer and sends $\hat{\theta}_i^{t+1}$ back to the server. While $\hat{\theta}_i^{t+1}$ is expected to be a faithful version of $\theta_i^{t+1}$ (except for known transformations like compression) for honest clients, this may not be the case for malicious clients. The server computes the updated global model parameters using the DOS aggregation rule. The proposed DOS aggregation rule consists of three key steps:

\noindent \textbf{Distance Computation}: The server starts with calculating the Euclidean and cosine distances between the local parameters sent by the clients as follows:

\begin{equation*}
    d^E_{ij} = ||\hat{\theta}_i - \hat{\theta}_j||_{2},
\qquad 
   d^C_{ij} = 1 - \tfrac{\hat{\theta}_i^{T} \hat{\theta}_j}{||\hat{\theta}_i||_{2} ||\hat{\theta}_j||_{2}}, 
\end{equation*}

\noindent where $i,j = 1,2,\ldots,n$. Here, the time index $t$ is skipped for convenience. The $(n \times n)$ distance matrices $M_E = [d^E_{ij}]$ and $M_C = [d^C_{ij}]$ are then computed and utilized for outlier detection. Unlike existing methods such as \cite{guerraoui2018hidden,yin2018byzantine,blanchard2017machine}, where outlier detection is performed directly on the model parameter space, the proposed DOS method performs outlier detection in the distance space.

\noindent \textbf{Outlier Score Computation}: After exploring various anomaly detection methods such as Random Forest \cite{primartha2017anomaly}, Local Outlier Factor (LOF) \cite{cheng2019outlier}, and K-means \cite{likas2003global}, we selected the Copula Based Outlier Detection (COPOD) method \cite{li2020copod} due to several factors. Firstly, other methods require choosing the percentage of abnormal data points (in our case clients) in advance, whereas COPOD is parameter-free, thus making it more robust. Moreover, COPOD is known to be computationally efficient even in high dimensional settings because it utilizes the marginals of the joint distributions, thereby allowing for greater flexibility and individual modeling of each dimension. Finally, the COPOD method also has desirable interpretability properties since it is based on modeling the tail probabilities along each dimension. In general, the COPOD function takes a $(n \times d)$ matrix as input, where each row represents a sample and each column represents an attribute of the sample to produce a $n$-dimensional vector of outlier scores $\mathbf{r}=[r_1, r_2, \ldots, r_n]$, where $r_i \in (0,+\infty)$ represents the relative likelihood that the sample $i$ is an outlier (true outliers are expected to have larger values of $r_i$). In DOS, we compute $\mathbf{r}_E = COPOD(M_E)$ and $\mathbf{r}_C = COPOD(M_C)$ and average these two vectors to obtain the final outlier score $\mathbf{r} = (\mathbf{r}_E+\mathbf{r}_C)/2$.

\noindent \textbf{Weighted Average Aggregation}: Ideally, the local parameter updates of those clients with higher outlier scores must be suppressed and the local parameters of clients with lower outlier scores must be amplified. To achieve this goal, we apply a softmax function to the outlier score vector with a temperature parameter of $-1$ and use the resulting output as the normalized weights for each client, i.e.,
\begin{equation*}
    w_i = \tfrac{exp(-r_i)}{\sum_{j=1}^n exp(-r_i)},\quad i = 1,2,\cdots,n.
\end{equation*}

\noindent The local parameters of the clients are aggregated using the following rule:
\begin{equation*}
    \theta^{t+1} = \textstyle{\sum}_{i=1}^n w_i \hat{\theta}_i^{t+1}.
\end{equation*}

\begin{figure}[t]
\centering
  \includegraphics[width=0.42\linewidth]{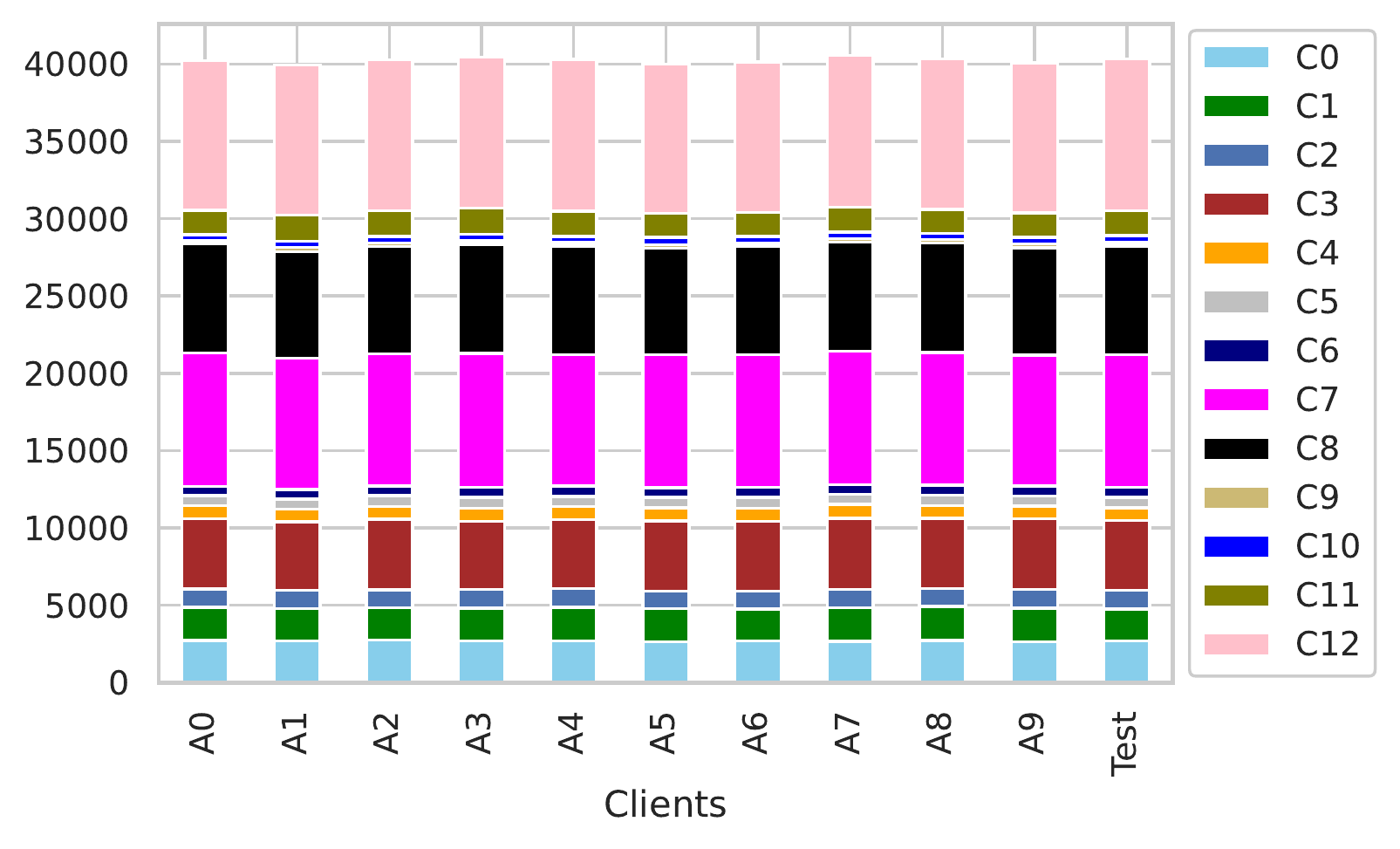}
$\qquad$
  \includegraphics[width=0.42\linewidth]{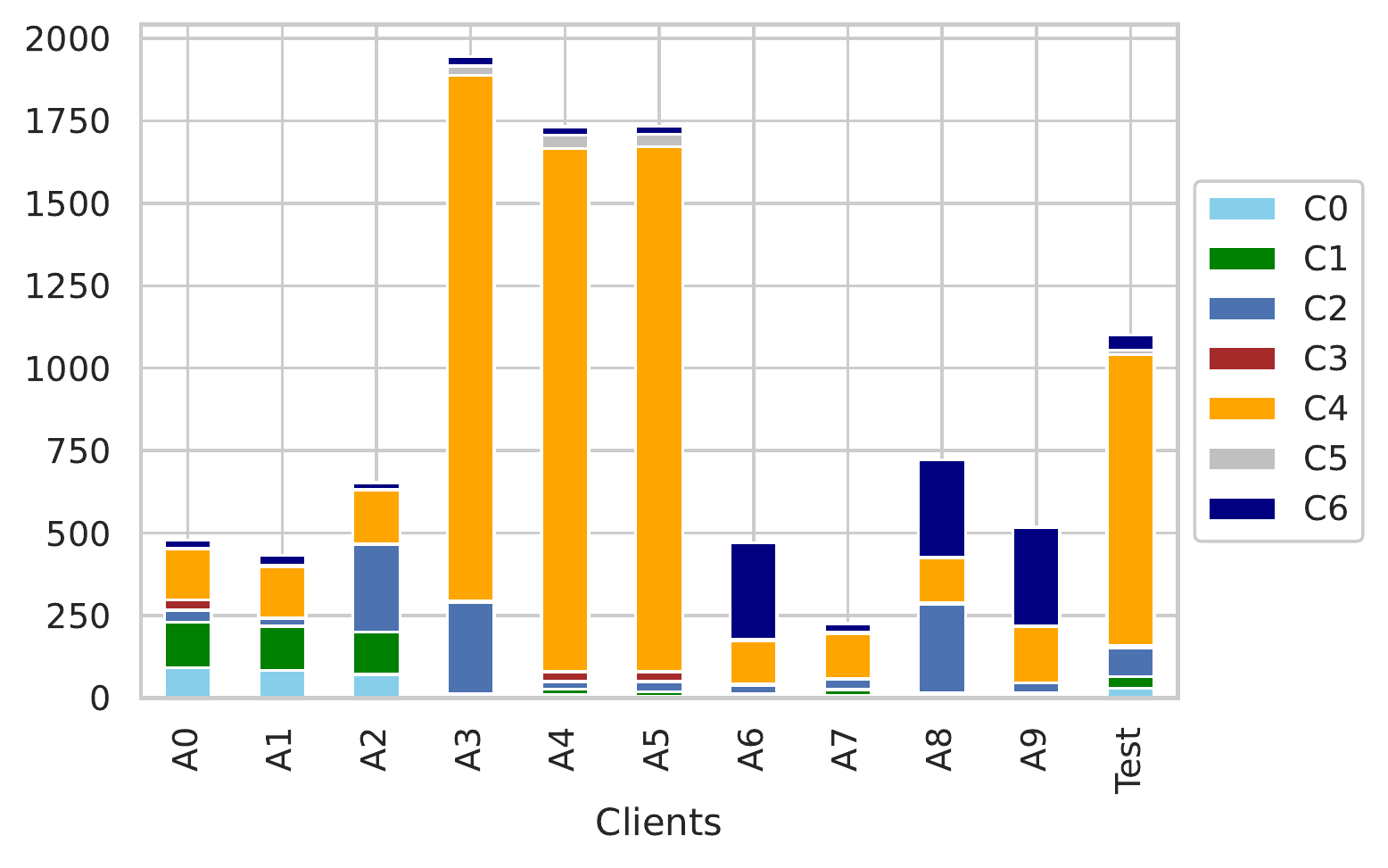}
\caption{{\bf Left:} Distribution of the multiclass multilabel CheXpert dataset among clients where each stacked bar represents the number of positive cases for class 0 to 12.
{\bf Right:} Distribution of the multiclass HAM1000 dataset in non-iid case among clients where each stacked bar represents the number of samples for class 0 to 6.
}\label{fig2}
\end{figure}

\section{Experiments}

\subsection{Datasets}

\textbf{CheXpert:} CheXpert is a large chest X-ray dataset with uncertainty labels and expert comparison \cite{irvin2019chexpert}. We use `CheXpert-small', a multi-class multi-label dataset that contains 191,456 chest X-ray images for training. The dataset is imbalanced and has 13 pathology categories. In addition, it has a percentage of uncertain data, and \cite{irvin2019chexpert} suggests either ignoring the uncertain labels during training or mapping all instances to zero or one. In our experiments, all uncertain labels were mapped to zero. The TorchXRayVision library \cite{Cohen2020xrv} was used to preprocess the data, where every image has a $224 \times 224$ resolution. We divide the training images equally between the clients as shown in Figure \ref{fig2} (\textbf{Left}), where the last stacked bar represents the testing dataset that we use to measure the global model's performance after each round. We use the CheXpert dataset to evaluate our method in the iid setting.

\noindent \textbf{HAM10000:} HAM10000 is a multi-class dataset consisting of 10,015 dermoscopic images from different populations \cite{tschandl2018ham10000}. It consists of 7 categories of pigmented lesions where every image was resized to $128 \times 128$. The train-test split is 8,910 and 1,105 images, respectively. We use this dataset to test our method in the non-iid case as shown in Figure \ref{fig2} (\textbf{Right}).

\begin{table}[t]
\centering
\caption{Area Under the Receiver Operating Characteristic Curve (AUC) with different types of poisoning attack scenarios on the Chexpert dataset. T-M stands for Trimmed Mean \cite{yin2018byzantine}. The last column represents the average AUC over five presented scenarios.}
\resizebox{1\textwidth}{!}{%

\begin{tabular}{ccccccc}
\toprule
{} &     No Attack &     Label Flip 10\% &     Mix Attack 40\% &     Noise 40\% & Noise \& Scaled 40\%  & Avg \\
\midrule

FedAvg &  0.70 &  0.69 &  0.50 &  0.50 &  0.50 & 0.58\\
Median &  0.70 &  0.69 &  0.69 &  0.69 &  0.54 & 0.66 \\
T-M &  0.70 &  0.69 &  0.69 &  0.50 &  0.50 & 0.62 \\
Krum &  0.66 &  0.66 &  0.64 &  0.63 &  0.67 & 0.65 \\
\textbf{DOS} &  \textbf{0.70} &  \textbf{0.69} &  0.68 &  \textbf{0.69} &  \textbf{0.67} & \textbf{0.68} \\
\bottomrule
\end{tabular}
}
\label{labcomp}
\end{table}

\subsection{FL Setup}

In all our experiments, we assume $n=10$ clients. The well-known ResNet-18 \cite{he2016deep} model, as well as a custom Convolutional Neural Network (CNN) with two convolutional layers, a ReLU activation function, and two fully connected layers are used as the  architectures for both global and local models. Each client only has access to its local dataset and the local models are trained using Stochastic Gradient Descent (SGD) \cite{gardner1984learning} with a learning rate of $0.01$. We implement all our experiments using the Pytorch framework \cite{NEURIPS2019_9015}. Using a batch size of 16 for CheXpert, each client is trained for a total of 100 rounds with 1 local step in each round that goes through the entire dataset. As for the HAM10000 dataset, we iterate through the whole dataset with a batch size of 890 for each client. We train it for a total of 250 rounds with 5 local steps for each client. This experiment was performed to show that the batch size does not impact the DOS aggregation rule. The evaluation metric is the Area Under the Receiver operating characteristic curve (AUC), which is calculated after each round of communication on the testing dataset. For the CheXpert dataset \cite{irvin2019chexpert}, the AUC was calculated for each class and the macro average was taken over all classes. For the HAM1000 dataset \cite{tschandl2018ham10000}, the average AUC was computed for all possible pairwise class combinations to minimize the effect of imbalance \cite{hand2001simple}.

\begin{figure}[t]
\centerline{{\includegraphics[scale = 0.15]{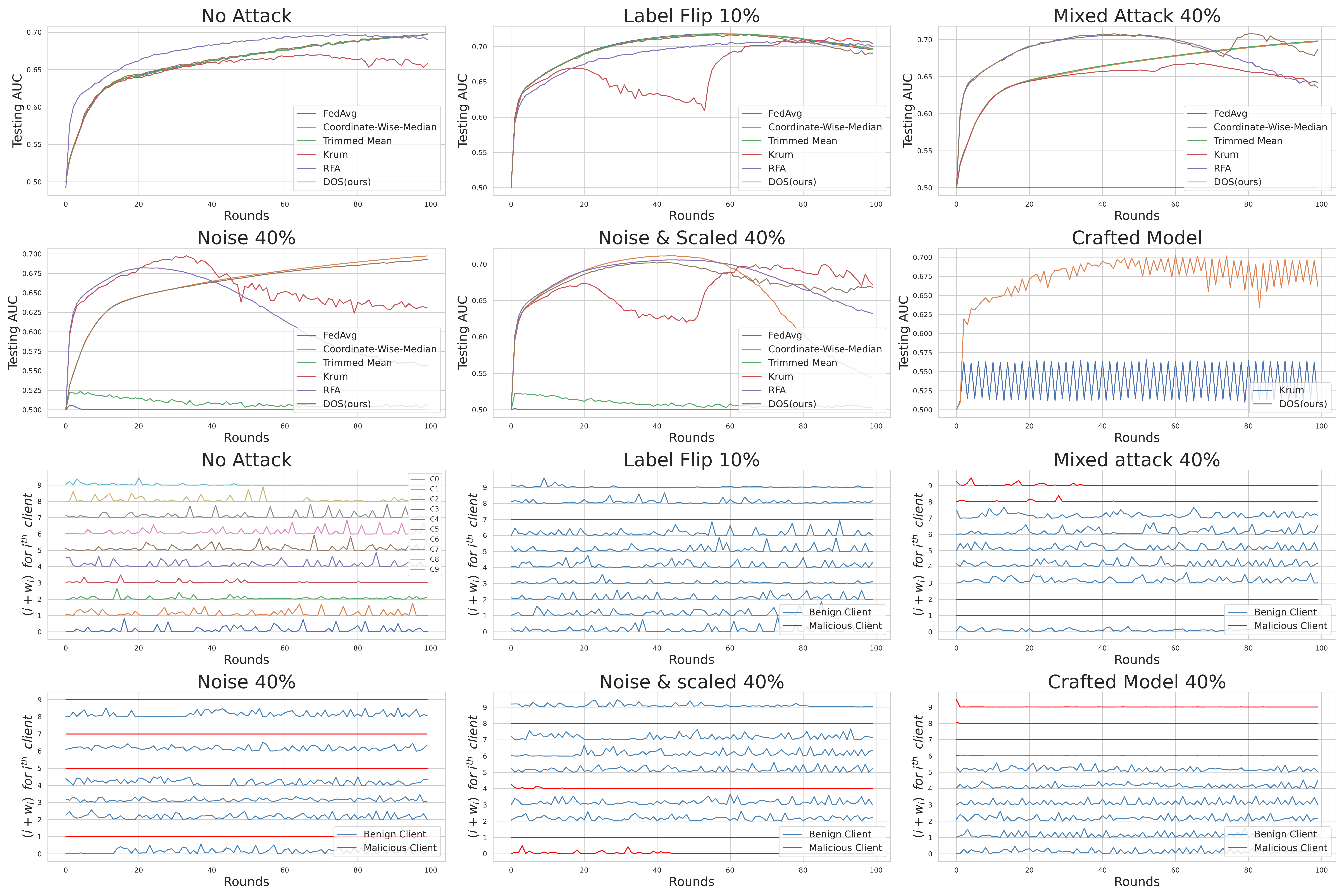}}}
\caption{Performance on CheXpert dataset using ResNet-18 model: The first two rows show AUC on a test set after each round for six scenarios in the following order (left to right): No Attack, Label Flip $10$\%, Mix Attack $40$\%, Noise $40$\%, Noise and Scaled $40$\%, and Crafted Attack $40$\%. The bottom two rows show the normalized weights of each client after each round.}  \label{fig3}
\vskip-11pt
\end{figure}

\subsection{Poisoning Attacks and Baseline Aggregation Rules}

We assume that up to $p \leq 40$\% of these clients could be malicious, i.e., at most 4 out of the 10 clients are malicious. We consider 5 different attacks on the CheXpert dataset: (i) \textbf{Label Flip $10$\%} - label-flipping by one of the clients, (ii) \textbf{Mix Attack $40$\%} - transmission of Gaussian noise by two clients and label-flipping by two clients, (iii) \textbf{Noise~$40$\%} - transmission of Gaussian noise by four clients, (iv) \textbf{Noise and Scaled $40$\%} - transmission of a mix of Gaussian noise and scaled parameters by four clients, and (v) \textbf{Crafted~$40$\%} - crafted model attack with four clients based on \cite{fang2020local} (with the aggregation rule being unknown to the attacker). On the HAM1000 dataset, 2 attacks were considered: (i) \textbf{Noise~$30$\%} - transmission of Gaussian noise by three clients, and (ii) \textbf{Mix Attack~$40$\%} - transmission of Gaussian noise by two clients, scaled parameter by a factor of 100 by one client, and scaled parameter by a factor of $-0.5$ (directionally opposite to the true parameters) by one client. The robustness of the proposed DOS method was benchmarked against the following aggregation rules: FedAvg \cite{mcmahan2017communication}, Coordinate-Wise-Median \cite{yin2018byzantine}, Trimmed Mean \cite{yin2018byzantine} and Krum \cite{blanchard2017machine}.  For both datasets, a \textbf{No Attack} scenario was also considered to evaluate the performance of DOS when all the clients are benign (honest). 

We conducted experiments by fixing the number of clients and increasing the proportion of malicious clients from 10\% to 60\%, in steps of 10\%. As expected, the DOS approach was robust until the proportion of malicious clients was less than or equal to 50\% and failed when the proportion was 60\% (e.g., for HAM10000 dataset, the AUC values were 0.695, 0.697, 0.696, 0.711, 0.710, and 0.554 for 10\%, 20\%, 30\%, 40\%, 50\%, and 60\% corruption, respectively. In comparison, the AUC without any attack was 0.70). This is the reason for choosing the proportion of malicious clients as 40\% for most of our experiments. Also, we conducted experiments by fixing the proportion of malicious clients to 40\% and increasing the number of clients from 5 to 40. The AUC values were 0.725, 0.700, 0.692, and 0.674 for 5, 10, 20 and 40 clients, respectively. We observe a minor degradation in accuracy when the number of clients increases, that it requires more rounds to converge.

\begin{figure}[t]
\centerline{
{\includegraphics[scale = 0.15]{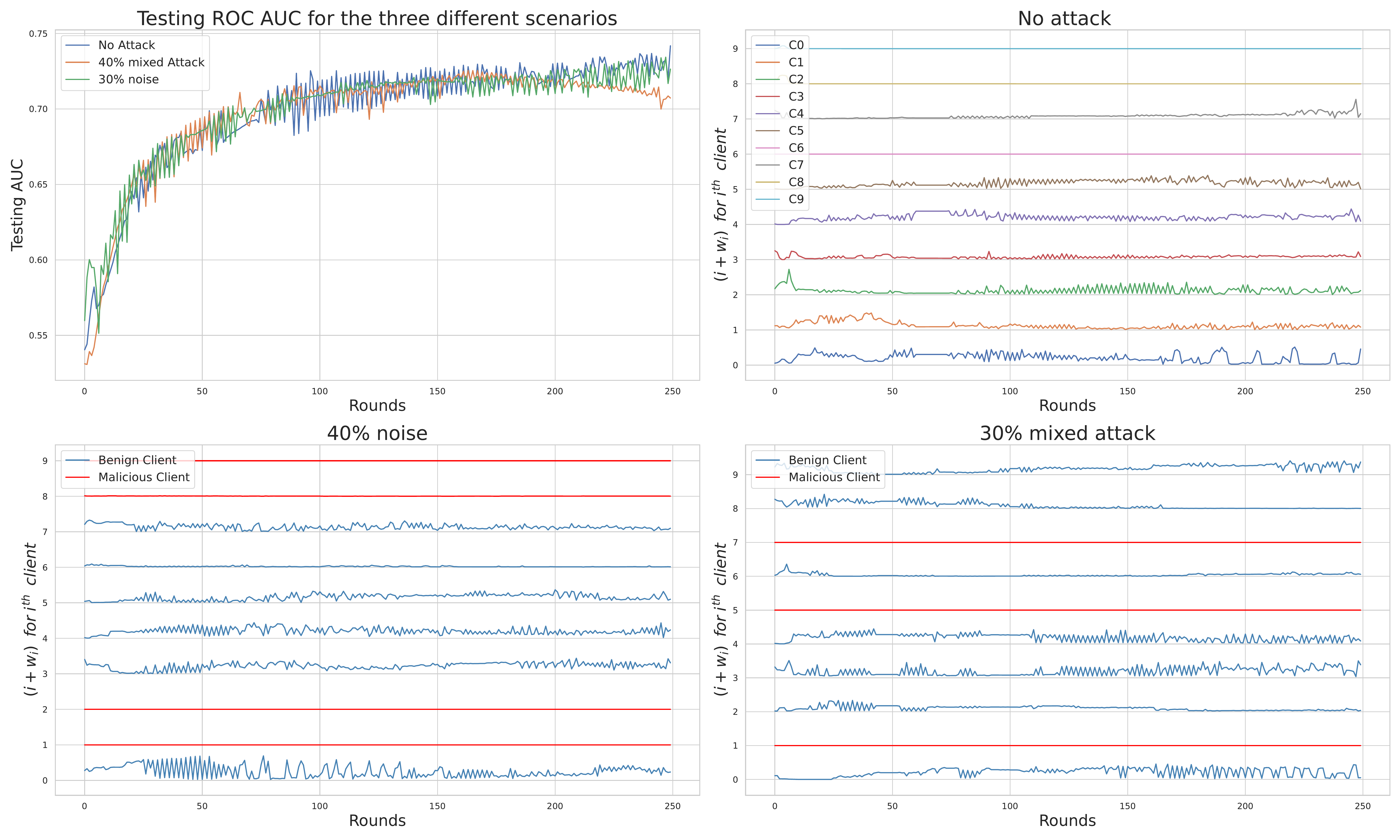}}}
\caption{Performance on HAM10000 dataset using a custom CNN model. The top-left plot shows AUC on the test set for DOS aggregation for the 3 scenarios.The normalized weights of each client after each round for the 3 scenarios are also depicted.} \label{fig4}
\end{figure}

\subsection{Results and Discussion}

Figure \ref{fig3} compares all the aggregation rules under six different scenarios on the CheXpert dataset. In the No Attack case, the DOS method outperforms Krum and produces comparable results to the other methods including FedAvg, which was essentially designed for benign settings. When one of the clients was trained by flipping class label 0 to 1, all of the methods performed well with Krum having a marginal advantage. For the mixed attack case, DOS had a faster convergence rate in the beginning with a marginal drop in round 70 whereas FedAvg was compromised. When four clients transmitted random Gaussian noise, FedAvg and Trimmed Mean were compromised, whereas DOS and Coordinate-Wise-Median performed best. In the noise and scaled parameters attack, FedAvg and Trimmed Mean were compromised from the start while the Coordinate-Wise-Median's performance started deteriorating at round 60. The DOS method and Krum performed best in this scenario. In the final crafted model setting, we compared DOS only with Krum as this attack was specifcially designed against Krum. In this attack, we can clearly see that DOS significantly outperforms Krum with a testing AUC of 0.65 vs. 0.50. Table \ref{labcomp} summarizes the average AUC for all rounds and it can be observed that DOS aggregation rule performs consistently well against all attack scenarios. In contrast, all the other aggregation rules had lower accuracy values in one or more scenarios. Furthermore, it can be observed from Figure \ref{fig3} that the weights of the malicious clients (shown in red) are almost always close to zero, indicating that the DOS method is able to effectively suppress the parameter updates from malicious clients.

Figure \ref{fig4} shows the performance of the DOS rule under three different scenarios on the HAM10000 dataset. In all cases, DOS performed with a high AUC score without dropping its overall performance. Similar to the CheXpert dataset, the weights of malicious clients are mostly negligible, demonstrating the effectiveness of the DOS method. The main advantage of DOS lies in its ability to account for both Euclidean and cosine distance, thereby addressing both positive and negative scaling. It also successfully detects Gaussian noise attacks and label-flip attacks, leveraging the strengths of the COPOD method. Since DOS combines three different approaches, it is hard to design model poisoning attacks that can successfully circumvent it. 

\subsection{Conclusions and Future Work}

In this paper, we present Distance-based Outlier Suppression (DOS), a novel robust aggregation rule that performs outlier detection in the distance space to effectively defend against Byzantine failures in FL. We prove the effectiveness and robustness of our method using two real-world medical imaging datasets. In future, we aim to extend this framework to ensure fairness between FL clients, based on the contribution of their local datasets. 

\bibliography{mybibliography}
\bibliographystyle{splncs04}

\end{document}